\documentclass[a4paper,10pt]{article}
\usepackage{eurosym}
\usepackage[dvips]{graphicx}
\usepackage{amsfonts}
\usepackage{mathrsfs}
\usepackage{graphicx} 
\usepackage{epsfig}
\usepackage{amsmath, amsthm}
\usepackage{amssymb}

\newcommand{\ba}{\begin{array}}
\newcommand{\ea}{\end{array}}

\newcommand{\beq}{\begin{equation}}
\newcommand{\eeq}{\end{equation}}
\newcommand{\ben}{\begin{enumerate}}
\newcommand{\een}{\end{enumerate}}
\newcommand{\bit}{\begin{itemize}}
\newcommand{\eit}{\end{itemize}}

\begin{document}
\title{Zakharov-Ito equation  and Generalized Heisenberg ferromagnet-type equation: equivalence and related   geometric curve flows}
\author{Zhanbala Umbetova\footnote{Email: zumurzakhova@gmail.com}, \, Shynaray Myrzakul\footnote{Email: srmyrzakul@gmail.com}, \,   Kuralay Yesmakhanova\footnote{Email: kryesmakhanova@gmail.com}, \\Tolkynay Myrzakul\footnote{Email: kryesmakhanova@gmail.com},  Gulgassyl Nugmanova\footnote{Email: gnnugmanova@gmail.com} \, and Ratbay Myrzakulov\footnote{Email: rmyrzakulov@gmail.com}\\
\textsl{Eurasian International Center for Theoretical Physics and} \\ { Department of General \& Theoretical Physics}, \\ Eurasian National University,
Nur-Sultan, 010008, Kazakhstan
}
\date{}
\maketitle

\begin{abstract}
These results continue our  studies of integrable generalized Heisenberg ferromagnet-type equations (GHFE) and their equivalent counterparts. 
We consider the GHFE which is the spin equivalent of the Zakharov-Ito equation (ZIE). We have established that these equations are gauge and geometrical equivalent to each other. The integrable motion of the space curve induced by the ZIE is constructed. The 1-soliton solution of the GHFE is obtained from the seed solution of the ZIE.
\end{abstract}


\section{Introduction}
The famous Korteweg-de Vries (KdV) equation 
\begin{eqnarray}
u_{t}+6uu_{x}+u_{xxx}=0 \label{1}
\end{eqnarray}
is the first integrable equation in soliton theory. At present, there are exist several integrable and nonintegrable generalizations of the KdV equation in 1+1 and 2+1 dimensions.  For example, the Zakharov-Ito equation (ZIE) 
\begin{eqnarray}
u_{t}+6uu_{x}+u_{xxx}+0.5\rho\rho_{x}&=&0,  \label{2}\\
\rho_{t}+2(u\rho)_{x}&=&0 \label{3}
\end{eqnarray}
is one of such integrable generalizations of the KdV equation in 1+1 dimension.  Another interesting subclass of integrable systems is the  Heisenberg ferromagnet type equations. The pioneering example of this subclass is the  Heisenberg ferromagmet equation (HFE)
\begin{eqnarray}
iA_{t}+\frac{1}{2}[A,A_{xx}]=0, \label{4}
\end{eqnarray}
where
\begin{eqnarray}
A=\left(\ba{cc}A_{3}&A^{-}\\A^{+}&-A_{3}\ea\right), \quad A^{\pm}=A_{1}\pm iA_{2}, \quad A^{2}=I, \quad {\bf A}=(A_{1},A_{2},A_{3}). \label{5}
\end{eqnarray}
It is well-known that the HFE (\ref{4}) is gauge/geometrical equivalent to the nonlinear Schr\"odinger equation (NLSE) \cite{l1}-\cite{l2}
\begin{eqnarray}
iq_{t}+q_{xx}+2|q|^{2}q=0, \label{6}
\end{eqnarray}
where $q(x,t)$ is a complex function.
The purpose of  this paper is to find and  study  the GHFE which is gauge/geometrical equivalent counterpart of the ZIE. 

This paper is organized as follows. In Section 2, we give the GHFE,    its Lax representation (LR) and a reduction.  Geometric formulation of the ZIE  in terms of space  curves in 3-dimensional Euclidean space $R^{3}$ is presented in Section 3. Some detail informations of ZIE we give in Section 4. In Section 5, we have estalished the gauge equivalence between the ZIE and the GHFE. The 1-soliton solutions of the GHFE is presented in Section 6. Last section is devoted to the conclusion. 

\section{The Kuralay-I  equation}
There are several integrable and nonintegrable GHFE (see, e.g., \cite{1907.10910}-\cite{1301.0180}). One of the  representatives of such GHFE is the Kuralay-I equation (K-IE). In this section, we present some main informations of the K-IE.
\subsection{Equation}
The Kuralay-I equation  (K-IE)   has the form
\begin{eqnarray}
A_{t}+A_{xxx}+(B A)_{x}+\frac{1}{8\beta^{2}}[(\rho^{2})_{x}Z]_{x}-8i\beta u_{x}Z&=&0, \label{7}\\
Z_{t}-v_{21}^{'}A-2v_{11}^{'}Z&=&0, \label{8}
\end{eqnarray}
where  
\begin{eqnarray}
A&=&\left(\ba{cc}A_{3}&A^{-}\\A^{+}&-A_{3}\ea\right), \quad A^{2}= I, \quad {\bf A}=(A_{1},A_{2},A_{3}), \quad {\bf A}^{2}=1, \label{9}\\
Z&=&\left(\ba{cc}z_{11}&z_{12}\\z_{21}&-z_{11}\ea\right), \quad Z^{2}=0, \quad Z_{x}^{2}=I, \quad Z_{t}^{2}=v_{21}^{'2}I, \label{10}\\
B&=&A_{x}^{2}-(4\beta^{2}-2u)I+2i\beta A_{xx}. \label{11}
\end{eqnarray}
Here
\begin{eqnarray}
A_{x}^{2}&=&4(u-\frac{\rho^{2}}{16\beta^{2}})I, \quad Z_{x}^{2}=u_{21}^{'2}I,\label{12} \\
u&=&2\beta\mp  0.25tr(Z_{t}^{2}), \quad \rho^{2}=4\beta^{2}[8\beta\mp tr(Z_{t}^{2})-tr(A_{x}^{2}), \label{13}\\
Z_{x}&=&u_{21}^{'}A-2u_{11}^{'}Z, \label{14}\\
v_{11}^{'}&=&2i\beta u -u_{x}, \label{15}\\
v_{21}^{'}&=&-4\beta^{2}+2u, \label{16}\\
u_{11}^{'}&=&-i\beta, \label{17}\\
u_{21}^{'}&=&-1. \label{18}
\end{eqnarray}

\subsection{Lax representation}
The K-IE  (\ref{7})-(\ref{8}) is integrable. Its LR is given by
\begin{eqnarray}
\Phi_x&=&U_{1}\Phi, \label{19}\\
\Phi_t&=&V_{1}\Phi. \label{20}
\end{eqnarray}
Here
\begin{eqnarray}
U_{1}&=&-i(\zeta-\beta)A-\frac{\rho^{2}}{16}(\zeta^{-2}-\beta^{-2})Z, \label{21}\\
V_{1}&=&w A+(\zeta^{2}-\beta^{2})\{[A,A_{x}]+\frac{\rho^{2}}{4\beta^{2}}Z\}
+i(\zeta-\beta)\{A_{xx}+\frac{(\rho^{2})_{x}}{8\beta^{2}}Z+ \nonumber \\
&+&4[u-\frac{\rho^{2}}{16\beta^{2}}]A\}+\frac{u\rho^{2}}{8}(\zeta^{-2}-\beta^{-2})Z, \label{22}
\end{eqnarray}
where  
\begin{eqnarray}
  w=-4i(\zeta^{3}-\beta^{3})+2i(\zeta-\beta)u. \label{23}
\end{eqnarray}

\subsection{Reduction}
Note that if $\beta=0=\rho$, the K-IE equation admits the following reduction
\begin{eqnarray}
A_{t}+A_{xxx}+[(A_{x}^{2}+2uI) A]_{x}=0 \label{24}
\end{eqnarray}
or
\begin{eqnarray}
A_{t}+A_{xxx}+1.5(A_{x}^{2} A)_{x}=0. \label{25}
\end{eqnarray}

\section{Integrable motion of  space curves induced by the ZIE}
In this section, we want to  present the geometric formulation of the ZIE. To do this, let us consider a   space curve in $R^{3}$ with the position vector ${\bf \gamma}(x,t)$. Then ${\bf v}=\frac{d{\bf \gamma}}{dx}$ and ${\bf a}=\frac{d^{2}{\bf \gamma}}{dx^{2}}$ are the velocity and the acceleration. In this paper, we shall assume that $x$ is the natural parameter, i.e. the length along the curve. Thus, $x$ is the  arc length of the curve at each time $t$. In other words, the velocity has unit length: ${\bf v}^{2}=|{\bf v}|^{2}=1$. Then the acceleration vector is orthogonal to the velocity vector ${\bf v}\cdot {\bf a}=0$. The magnitude of the acceleration vector $\kappa=|{\bf a}|$ is called the curvature of the curve. Let us introduce three vectors: ${\bf e}_{1}\equiv{\bf v}, \quad {\bf e}_{2}=\frac{{\bf a}}{|{\bf a}|}, \quad {\bf e}_{3}={\bf e}_{1}\wedge {\bf e}_{2}$. From the differential geometry follows that such space curve ${\bf \gamma}(x,t)$  describes by the     Frenet-Serret equation (FSE). In this paper, we consider the more general form of the FSE with  two curvatures $\kappa_{1}\equiv\kappa$ and $\kappa_{2}$. Then the corresponding  FSE and its temporal counterpart are given by 
\begin{eqnarray}
\left ( \begin{array}{ccc}
{\bf  e}_{1} \\
{\bf  e}_{2} \\
{\bf  e}_{3}
\end{array} \right)_{x} = C
\left ( \begin{array}{ccc}
{\bf  e}_{1} \\
{\bf  e}_{2} \\
{\bf  e}_{3}
\end{array} \right),\quad
\left ( \begin{array}{ccc}
{\bf  e}_{1} \\
{\bf  e}_{2} \\
{\bf  e}_{3}
\end{array} \right)_{t} = G
\left ( \begin{array}{ccc}
{\bf  e}_{1} \\
{\bf  e}_{2} \\
{\bf  e}_{3}
\end{array} \right), \label{26} 
\end{eqnarray}
where  ${\bf e}_{1}={\bf \gamma}_{x}$ (the   unit tangent vector), ${\bf e}_{2}=\frac{{\bf \gamma}_{xx}}{|{\bf \gamma}_{xx}|}$ (principal normal vector) and ${\bf e}_{3}={\bf e}_{1}\wedge {\bf e}_{2}$ (the binormal vector), 
respectively.
Here
\begin{eqnarray}
C&=&
\left ( \begin{array}{ccc}
0   & \kappa_{1}     & \kappa_{2}  \\
-\kappa_{1}  & 0     & \tau  \\
-\kappa_{2}    & -\tau & 0
\end{array} \right),\quad G=\left ( \begin{array}{ccc}
0       & \omega_{3}  & \omega_{2} \\
-\omega_{3} & 0      & \omega_{1} \\
-\omega_{2}  & -\omega_{1} & 0
\end{array} \right)\label{27} 
\end{eqnarray}
or
\begin{eqnarray}
C=-\tau L_{1}+\kappa_{2}L_{2}-\kappa_{1}L_{3},  \quad G=-\omega_{1}L_{1}+\omega_{2}L_{2}-\omega_{3}L_{3},\label{28} 
\end{eqnarray}
where  $\tau$,  $\kappa_{1}, \kappa_{2}$ are the  "torsion",  "geodesic curvature" and  "normal curvature" of the curve, respectively; $\omega_{j}$ are some  functions. The basis elements of $so(3)$ are given by  
\begin{eqnarray}
L_{1}=
\left ( \begin{array}{ccc}
0   & 0     & 0  \\
0  & 0     & -1  \\
0    & 1 & 0
\end{array} \right), \quad L_{2}=
\left ( \begin{array}{ccc}
0       & 0  & 1 \\
0 & 0      & 0 \\
-1  & 0 & 0
\end{array} \right), \quad L_{3}=
\left ( \begin{array}{ccc}
0       & -1  & 0\\
1 & 0      & 0 \\
0  & 0 & 0
\end{array} \right). \label{29}
\end{eqnarray}
 They obey  the  commutation relations
\begin{eqnarray}
[L_{1}, L_{2}]=L_{3}, \quad [L_{2}, L_{3}]=L_{1},  \quad [L_{3}, L_{1}]= L_{2}.\label{30}
\end{eqnarray}
The   basis elements of $su(2)$ algebra read as 
\begin{eqnarray}
e_{1}=
\frac{1}{2i}\left ( \begin{array}{cc}
0       & 1  \\
1 & 0
\end{array} \right), \quad e_{2}=
\frac{1}{2i}\left ( \begin{array}{cc}
0       & -i  \\
i & 0
\end{array} \right), \quad e_{3}=
\frac{1}{2i}\left ( \begin{array}{cc}
1      & 0  \\
0 & -1
\end{array} \right), \label{31}
\end{eqnarray}
where 
\begin{eqnarray}
\sigma_{1}=
\left ( \begin{array}{cc}
0       & 1  \\
1 & 0
\end{array} \right), \quad \sigma_{2}=
\left ( \begin{array}{cc}
0       & -i  \\
i & 0
\end{array} \right), \quad \sigma_{3}=
\left ( \begin{array}{cc}
1      & 0  \\
 0& -1
\end{array} \right), \label{32}
\end{eqnarray}
are the Pauli matrices. These elements obey the  commutation relations:
\begin{eqnarray}
[e_{1}, e_{2}]=e_{3}, \quad [e_{2}, e_{3}]=e_{1},  \quad [e_{3}, e_{1}]= e_{2}. \label{33}
\end{eqnarray}
Note that the Pauli matrices obey the  commutation relations:
\begin{eqnarray}
[\sigma_{1}, \sigma_{2}]=2i\sigma_{3}, \quad [\sigma_{2}, \sigma_{3}]=2i\sigma_{1},  \quad [\sigma_{3}, \sigma_{1}]= 2i\sigma_{2}\label{34}
\end{eqnarray}
or
\begin{eqnarray}
[\sigma_{i}, \sigma_{j}]=2i\epsilon_{ijk}\sigma_{k}.\label{35}
\end{eqnarray}
The   isomorphism between the Lie algebras $su(2)$ and $so(3)$ induced   the following  correspondence between their basis elements $L_{j}\leftrightarrow e_{j}$. The compatibility condition of the equations (\ref{26}) has the form
\begin{eqnarray}
C_t - G_x + [C, G] = 0.\label{36} 
\end{eqnarray}
This matrix equation in terms of  elements  takes the form 
 \begin{eqnarray}
\kappa_{1t}- \omega_{3x} -\kappa_{2}\omega_{1}+ \tau \omega_2&=&0, \label{37} \\ 
\kappa_{2t}- \omega_{2x} +\kappa_{1}\omega_{1}- \tau \omega_3&=&0, \label{38} \\
\tau_{t}  -    \omega_{1x} - \kappa_{1}\omega_2+\kappa_{2}\omega_{3}&=&0.  \label{39} \end{eqnarray}
Let us  we assume that
\begin{eqnarray}
\kappa_{1}&=&-2iu_{11}, \quad \kappa_{2}=-(u_{12}-u_{21}), \quad \tau=-i(u_{12}+u_{21}),\label{40} \\
\omega_{1}&=&-i(v_{12}+v_{21}), \quad \omega_{2}=-(v_{12}-v_{21}), \quad \omega_{3}=-2iv_{11}, \label{41}
\end{eqnarray}
where
\begin{eqnarray}
u_{11} & = &-i\zeta, \quad u_{12}=u-\frac{\rho^{2}}{16\zeta^{2}}, \quad u_{21}=-1,\label{42}\\ 
v_{11}&=&-(4i\zeta^{3}-2i\zeta u+u_{x}), \label{43}\\
v_{12}&=&4\zeta^{2}u+2i\zeta u_{x}-(u_{xx}+2u^{2}+0.25\rho^{2})+\frac{u\rho^{2}}{8\zeta^{2}},      \label{44}\\
v_{21}&=&-(4\zeta^{2}-2u). \label{45}
\end{eqnarray}
Then  Eqs.(\ref{37})-(\ref{39}) give us the following equations for $u, \rho$:
\begin{eqnarray}
u_{t}+6uu_{x}+u_{xxx}+0.5\rho\rho_{x}&=&0, \label{46}\\
\rho_{t}+2(u\rho)_{x}&=&0. \label{47}
\end{eqnarray}
It is nothing but the ZIE. Thus we have constructed the integrable motion of the space curve induced by the ZIE.  Our next aim is to find the integrable spin system,  or  in other words, the integrable generalized Heisenberg ferromagnet type equation which is geometrical equivalent counterpart of the ZIE. To construct this spin systems, we introduce two vectors as \begin{eqnarray}
{\bf e}_{3} = (A_{1},A_{2},A_{3})\equiv{\bf A}, \quad {\bf Z}=(z_{1}, z_{2}, z_{3})=p_{1}{\bf e}_{1}+p_{2}{\bf e}_{2}+p_{3}{\bf e}_{3}, \label{48}
\end{eqnarray}
where ${\bf e}_{3}$ and ${\bf Z}$ are the vector equivalents of the matrix functions $A$ and $Z$, respectively. Here $p_{j}=p_{j}(x,t)$ are some functions to be determined and
\begin{eqnarray}
z_{1}=0.5(z_{21}+z_{12}), \quad z_{2}=-0.5i(z_{21}-z_{12}), \quad z_{3}=z_{11}. \label{49}
\end{eqnarray}
In terms of the functions $g_{j}$,  the components of the vectors ${\bf e}_{3}$ and  ${\bf Z}$ take the forms
\begin{eqnarray}
A_{1}&=&-\frac{g_{1}g_{2}+\bar{g}_{1}\bar{g}_{2}}{\Delta}, \quad  
A_{2}=-\frac{g_{1}g_{2}-\bar{g}_{1}\bar{g}_{2}}{i\Delta}, \quad   
A_{3}=\frac{|g|^{2}_{1}-|g|^{2}_{2}}{\Delta}, \label{50}\\
z_{1}&=&\frac{\bar{g}_{1}^{2}-g_{2}^{2}}{2\Delta}, \quad  z_{2}= \frac{\bar{g}_{1}^{2}+g_{2}^{2}}{2i\Delta}, \quad   
z_{3}=\frac{\bar{g}_{1}g_{2}}{\Delta}.\label{51}
\end{eqnarray}
We now ready to write the following set of equations
\begin{eqnarray}
{\bf e}_{3t} & = &-\omega_{2}{\bf e}_{1}-\omega_{1}{\bf e}_{2},\label{52}\\ 
{\bf Z}_{t}&=&b_{1}{\bf e}_{1}+b_{2}{\bf e}_{2}+b_{3}{\bf e}_{3}, \label{53}
\end{eqnarray}
where 
\begin{eqnarray}
b_{1}&=&p_{1t}-\omega_{3}p_{2}-\omega_{2}p_{3}, \nonumber\\
b_{2}&=&p_{2t}+\omega_{3}p_{1}-\omega _{1}p_{3}, \label{54}\\
b_{3}&=&p_{3t}+\omega_{2}p_{1}+\omega_{1} p_{2} \nonumber
\end{eqnarray}
are some real functions. In fact, this set of equations (\ref{52})-(\ref{53}) is the K-IE written in the slightly other (vector) form. This vector K-IE is equivalent to the matrix K-IE (\ref{7})-(\ref{8}). Thus  we have  proved,      the  Lakshmanan (geometrical)  equivalence between the ZIE and the K-IE.

\section{Zakharov-Ito equation}
For our convenience,  let us here one more present the ZIE (see, e.g., \cite{0906.0780})
\begin{eqnarray}
u_{t}+6uu_{x}+u_{xxx}+0.5\rho\rho_{x}&=&0, \label{55}\\
\rho_{t}+2(u\rho)_{x}&=&0. \label{56}
\end{eqnarray}
The ZIE is integrable. Its LR reads as
\begin{eqnarray}
\psi_{2xx}&=&-(\zeta^{2}+u-\frac{\rho^{2}}{16\zeta^{2}})\psi_{2}, \label{57}\\
\psi_{2t}&=&u_{x}\psi_{2}+(4\zeta^{2}-2u)\psi_{2x}. \label{58}
\end{eqnarray}
Note that as $\rho=0$, the ZIE admits the following  reduction
 \begin{eqnarray}
u_{t}+6uu_{x}+u_{xxx}=0, \label{59}
\end{eqnarray}
which is the famous KdV equation.

\section{Gauge equivalence between the ZIE and the K-IE}
In the section 3, we have proved the geometrical (Lakshmanan) equivalence between the K-IE and the ZIE. In this section we want to show that between the K-IE (\ref{7})-(\ref{8}) and the ZIE (\ref{55})-(\ref{56}) take place also the  gauge equivalence. To prove this statement,  let us consider the gauge transformation
\begin{eqnarray}
\Psi=g\Phi,\label{60}
\end{eqnarray}
where
\begin{eqnarray}
\Psi=\left(\ba{c}\psi_{1}\\
\psi_{2}\ea\right),\quad \psi_{1}=i\lambda\psi_{2}-\psi_{2x},\quad g=\Psi|_{\lambda=\beta}.\label{61}
\end{eqnarray}  Then this transformation induces the new Lax pair  $U_{3}-V_{3}$ from known $U_{1}-V_{1}$ (\ref{21})-(\ref{22}) as 
\begin{eqnarray}
U_{3}=gU_{1}g^{-1}+g_{x}g^{-1}, \quad V_{3}=gV_{1}g^{-1}+g_{t}g^{-1}.\label{62}
\end{eqnarray}
where
\begin{eqnarray}
U_{3}&=&\left(\ba{cc}-i\lambda& u-\frac{\rho^{2}}{16\lambda^{2}}\\ -1&i\lambda\ea\right)=-i\lambda\sigma_{3}+Q, \quad Q=\left(\ba{cc}0& u-\frac{\rho^{2}}{16\lambda^{2}}\\ -1& 0\ea\right), \label{63}\\
V_{3}&=&F_{3}\lambda^{3}+F_{2}\lambda^{2}+F_{1}\lambda+F_{0}+F_{-2}\lambda^{-2}=F-(4i\lambda^{3}-2i\lambda u+u_{x})\sigma_{3}.\label{64}
\end{eqnarray}
Here
\begin{eqnarray}
F_{3}&=&4i\sigma_{3}, \label{65}\\
F_{2}&=&4\left(\ba{cc}0&u\\
-1&0\ea\right), \label{66}\\
F_{1}&=&2iu\sigma_{3}+2i\left(\ba{cc}0&u_{x}\\
0&0\ea\right), \label{67}\\
F_{0}&=&-u_{x}\sigma_{3}+\left(\ba{cc}0&-(u_{xx}+\frac{\rho^{2}}{4}+2u^{2})\\
2u&0\ea\right), \label{68}\\
F_{-2}&=&\frac{u\rho^{2}}{8}\Sigma, \quad \Sigma=\left(\ba{cc}0&1\\
0&0\ea\right), \label{69}\\
F&=&\left(\ba{cc}0&4\lambda^{2}u+2i\lambda u_{x}- (u_{xx}+\frac{\rho^{2}}{4}+2u^{2})+\frac{u\rho^{2}}{8\lambda^{2}}\\
-(4\lambda^{2}-2u)&0\ea\right).\label{70}
\end{eqnarray}
Note that 
\begin{eqnarray}
V_{3}=4\lambda^{2} U_{3}+V^{'}_{3}=4\lambda^{2}U_{3}+F_{2}^{'}\lambda^{2}+F_{1}\lambda+F_{0}^{'}+F_{-2}\lambda^{-2}.\label{71}
\end{eqnarray}
We now consider the linear equations
\begin{eqnarray}
\Psi_{x}&=&U_{3}\Psi,\label{72}\\
\Psi_{t}&=&V_{3}\Psi.\label{73}
\end{eqnarray}
The compatibility condition of these linear equations
\begin{eqnarray}
U_{3t}-V_{3x}+[U_{3},V_{3}]=0 \label{74}
\end{eqnarray}
gives the ZIE (\ref{55}))-(\ref{56}). In fact, we have
\begin{eqnarray}
\lambda^{0}&:& u_{t}+6uu_{x}+u_{xxx}+0.5\rho\rho_{x}=0, \label{75}\\
\lambda^{-2}&:& \rho_{t}+2(u\rho)_{x}=0, \label{76}\\
\lambda^{j}&:& 0=0 \quad (j=1,2,3).\label{77}
\end{eqnarray}
In our case, the matrices $A$ and $Z$ have the following forms
\begin{eqnarray}
A=g^{-1}\sigma_{3}g, \quad Z=g^{1}\Sigma g, \quad \Sigma=\left(\ba{cc}0& 1\\
0&0\ea\right), \quad g=\left(\ba{cc}g_{1}&-\bar{g}_{2}\\
g_{2}&\bar{g}_{1}\ea\right) \label{78}
\end{eqnarray}
or
\begin{eqnarray}
A=\frac{1}{\Delta}\left(\ba{cc}|g_{1}|^{2}-|g_{2}|^{2}&-2\bar{g}_{1}\bar{g}_{2}\\
-2g_{1}g_{2}&|\bar{g}_{2}|^{2}-|g_{1}|^{2}\ea\right),\quad
 Z=\frac{1}{\Delta}\left(\ba{cc}\bar{g}_{1}g_{2}&\bar{g}_{1}^{2}\\
-g^{2}_{2}&-\bar{g}_{1}g_{2}\ea\right). \label{79}
\end{eqnarray}
From the gauge equivalence between the ZIE  and the K-IE follows the   some important relations between the solutions of these equations. Some of them look like as
\begin{eqnarray}
Z^{2}&=&0, \quad Z_{x}^{2}=I, \quad Z_{t}^{2}=v_{21}^{'2}I. \label{80}\\
A_{x}^{2}&=&4(u-\frac{\rho^{2}}{16\beta^{2}})I, \quad Z_{x}^{2}=u_{21}^{'2}I. \label{81} 
\end{eqnarray}

\section{Soliton  solutions of the K-IE}
As the integrable equation, the K-IE has all ingredients of integrable systems like LR, conservation laws, Hamiltonian structures, soliton solutions and so on.  Here let us present the 1-soliton  solution of the K-IE. To construct this 1-soliton solution, we use the  gauge equivalence between the K-IE and the ZIE. First recall that the matrices $A$ and $Z$ can be expressed as 
\begin{eqnarray}
A=g^{-1}\sigma_{3}g=
\left ( \begin{array}{cc}
A_{3}       & A^{-}  \\
A^{+} & -A_{3}
\end{array} \right),\quad  
Z=g^{-1}\Sigma_{3}g=
\left ( \begin{array}{cc}
z_{11}       & z_{12}  \\
z_{21} & -z_{11}
\end{array} \right), \label{82}\end{eqnarray}
where
\begin{eqnarray}
g=
\left ( \begin{array}{cc}
g_{1}      & -\bar{g}_{2} \\
g_{2} & \bar{g}_{1}
\end{array} \right), \quad g^{-1}=\frac{1}{\Delta}
\left ( \begin{array}{cc}
\bar{g}_{1}       & \bar{g}_{2}  \\
-g_{2} & g_{1}
\end{array} \right), \quad \Delta=|g_{1}|^{2}+|g_{2}|^{2}. \label{83}\end{eqnarray}
As result we obtain the following expressions  for the components of the matrices $A$ and $Z$:
\begin{eqnarray}
A^{+}&=&-\frac{2g_{1}g_{2}}{\Delta}, \quad A_{3}=\frac{|g_{1}|^{2}-|g_{2}|^{2}}{\Delta}, \label{84}\\
z_{11}&=&\frac{\bar{g}_{1}g_{2}}{\Delta}, \quad z_{12}=\frac{\bar{g}_{1}^{2}}{\Delta}, \quad z_{21}=-\frac{g_{2}^{2}}{\Delta}.  \label{85}
\end{eqnarray}
To construct the 1-soliton  solution of the K-IE, we consider the following seed solution of the ZIE
\begin{eqnarray}
u=\rho=0.  \label{86}
\end{eqnarray}
Then we obtain the following equations for $g_{j}$:
\begin{eqnarray}
g_{1x}&=&-i\beta g_{1},\label{87}\\
g_{2x}&=&-g_{1}+i\beta g_{2},\label{88}\\
g_{1t}&=&-4i\beta^{3}g_{1}, \label{89}\\
g_{2t}&=&-4\beta^{2}g_{1}+4i\beta^{3}g_{2}. \label{90}
\end{eqnarray}
These  equations  admit the following solution
\begin{eqnarray}
g_{1}=a_{1}e^{-\theta}, \quad g_{2}=a_{2}e^{\theta+\delta}, \label{91} 
\end{eqnarray}
where $a_{j}$ are complex constants and 
\begin{eqnarray}
\theta=i\beta x+4i\beta^{3}t, \quad \delta=-a_{1}a_{2}^{-1}(x+4t).  \label{92}
\end{eqnarray}
 Thus the 1-soliton solution of the K-IE has the form
\begin{eqnarray}
A^{+}&=&-\frac{e^{\delta-\bar{\delta}+i(\delta_{1}+\delta_{2})}}{\cosh\chi}, \quad A_{3}=\tanh\chi,\quad z_{11}= \frac{e^{(\theta-\bar{\theta})+0.5(\delta-\bar{\delta})+i(\delta_{2}-\delta_{1})}}{2\cosh\chi}, \label{93}\\
z_{12}&=&\frac{e^{-2\bar{\theta}-0.5(\delta+\bar{\delta})-2i\delta_{1}-\delta_{3}}}{2\cosh\chi},\quad z_{21}=-\frac{e^{2(\theta+\delta)-0.5(\delta+\bar{\delta})+2i\delta_{2}+\delta_{3}}}{2\cosh\chi},  \label{94}
\end{eqnarray}
where $\delta_{3}=\ln|\frac{a_{2}}{a_{1}}|,\quad a_{j}=|a_{j}|e^{i\delta_{j}}, \quad \chi=\theta+\bar{\theta}+0.5(\delta+\bar{\delta})$. Finally we note  that   the presented  1-soliton solution satisfies the following conditions 
\begin{eqnarray}
|A^{+}|^{2}+A_{3}^{2}=1, \quad z_{11}^{2}+z_{12}z_{21}=0, \label{95}
\end{eqnarray}
 which come  from the following properties of the matrices $A$ and $Z$:
 \begin{eqnarray}
A^{2}=I, \quad Z^{2}=0.\label{96}
\end{eqnarray}
\section{Conclusion}
In the paper, one of the integrable GHFE,  namely,  the K-IE  is investigated. The LR  of this equation and its reduction are given.  The geometric formulation of the ZIE  is presented. It is also shown that the ZIE  is geometrical and gauge equivalent to the K-IE. The 1-soliton solution of the K-IE is obtained. 
Finally we note that it is interesting to investigate the surfaces induced by the K-IE and by the ZIE as well as integrable motion of curves in  other geometries  (see, e.g., \cite{1907.10910}-\cite{1301.0180}). 

\section*{Acknowledgements}
This work was supported  by  the Ministry of Edication  and Science of Kazakhstan under
grants 0118РК00935 and 0118РК00693.

\end{document}